\documentstyle[12pt,aaspp,psfig]{article}
\slugcomment{\shortstack[r]{re-submitted to MNRAS, 19th June 2002}}
\begin{document}
\newcommand\Msun {M_{\odot}\ }
\newcommand\Lsun {L_{\odot}\ }
\newcommand\vsun {v_{\odot}\ }

\title{An Optical Velocity for the Phoenix Dwarf Galaxy\footnotemark[1]}
\vskip0.5cm

\author{{\bf Mike Irwin\footnotemark[2]}}
\affil{Institute of Astronomy, University of Cambridge, 
Madingley Road, Cambridge, CB3 0HA, England, UK}

\author{{\bf Eline Tolstoy\footnotemark[3]}}
\affil{Kapteyn Institute, University of Groningen, PO Box 800,
9700AV Groningen, the Netherlands}

\footnotetext[1]{Based on observations collected at the European Southern
Observatory, Chile, in service mode, proposal number 66.B-0712(A)}
\footnotetext[2]{email: mike@ast.cam.ac.uk}
\footnotetext[3]{email: etolstoy@astro.rug.nl}

\begin{abstract}

We present the results of a VLT observing program carried out in
service mode using FORS1 on ANTU in Long Slit mode to determine the
optical velocities of nearby low surface brightness galaxies.  As part
of our program of service observations we obtained long-slit spectra
of several members of the Phoenix dwarf galaxy from which we derive an
optical helio-centric radial velocity of $-$13~$\pm$~9~km/s.  This
agrees very well with the velocity of the most promising of the HI
clouds seen around Phoenix, which has a helio-centric velocity of $-$23
km/s, but is significantly different to the recently published optical
heliocentric velocity of Phoenix of $-$52~$\pm$~6~ km/s of Gallart {\it
et al.} (2001).

\end{abstract}

\keywords{GALAXIES: INDIVIDUAL: PHOENIX, GALAXIES: KINEMATICS AND DYNAMICS,  
GALAXIES: LOCAL GROUP}

\section{Introduction}

Dynamical measurements of outlying Local Group galaxies are crucial 
for constraining both the age and the total mass of the Local Group
and for probing the nature of the intrinsic dark matter within the galaxies.
A necessary part of this process involves investigating the link between 
possible HI detections, the optical components of the galaxies and
the relevance of the HI to the recent star formation history of the system. 
Optical velocities determine if observations of HI gas in and around these 
systems are the result of gas associated with these galaxies, a chance
superposition with high velocity HI clouds, outlying components of the 
Magellanic Stream, or just Galactic foreground contamination.

Phoenix is a member of the Local Group, lying about 450~kpc from our
Galaxy (Ortolani \& Gratton 1988) in a fairly isolated location on the
opposite side of the Galaxy from the Andromeda sub-system (see
Table~1).  It is possibly one of the most distant of the satellites
of our Galaxy, or perhaps a free-floating outlying Local Group object.
It appears to be a galaxy in transition between a dwarf irregular (dI)
and a dwarf spheroidal (dSph) system, having had both a recent burst
of star formation and a plausible detection of $\approx 10^5 \Msun$ of
HI.  The HI gas has been detected close to the position of Phoenix at
several different locations, and velocities ($\vsun =$ 56 km/s, 120
km/s Carignan, Demers \& C\^{o}t\'{e} 1991; $\vsun = -$23 km/s
Oosterloo, Da Costa \& Staveley-Smith 1996; 
and from mosaic mapping using the Australia
Telescope Compact Array at $\vsun = -$23, 7, 59, 140 km/s St-G\'{e}rmain
{\it et al.} 1999 ).  Due to the superior resolution/sensitivity we
adopt the latter measurements as defining the possible HI-associated
gas throughout the remainder of this paper.

The HI complex at $\vsun =$ 140~km/s is thought to be an outlying component
of the Magellanic Stream, which passes close to the line-of-sight of Phoenix,
while the component at $\vsun = -$7~km/s is undoubtedly Galactic in origin. 
This leaves two remaining HI components which may be associated with Phoenix.
The more compact of the HI clouds at $\vsun = -$23~km/s is centred about 5
arc-minutes to the southwest of the optical centre of Phoenix and overlaps
the optical image.  This is within a distance of about 650 pc if it lies at 
the same distance as Phoenix, and partially covers the optical image which 
is about 1 kpc in
size.  The component with $\vsun = $ 59~km/s is much more extended, fragments
into four substructures, and is located significantly further to the south of 
the optical centre.  From the HI morphology and dynamics while it is unclear 
whether or not the 59~km/s component is, or perhaps was in the past, 
associated with Phoenix, the evidence for the $-$23~km/s system is more 
compelling.  The derived HI mass of $\approx 10^5 \Msun$ is comparable to
that found in the Sculptor dwarf spheroidal and in the dwarf irregular/dwarf
spheroidal transition object LGS 3 (Young \& Lo 1997).   Furthermore, as
St. G\'{e}rmain {\it et al.} point out, the HI velocity field of the $-$23~km/s
component shows clear evidence of a velocity gradient indicating either 
rotation, or ejection, from Phoenix.

The resolved stellar population of Phoenix has been studied in some detail
(Ortolani \& Gratton 1988; van de Rydt, Demers, \& Kunkel 1991; 
Held, Saviane \& Momany 1999; Mart\'inez-Delgado, Gallart \& Aparicio 1999;
Holtzmann, Smith \& Grillmair 2000).
Phoenix has obviously experienced recent star formation as can be seen from the
sprinkling of blue stars across the field. They are concentrated in
``associations'' near the centre of the galaxy and elongated in the
direction of the HI cloud at $-$23km/s.
This recent ``episode'' of star formation has been
quantified by Held {\it et al.} (1999) to have started at least 0.6 Gyr ago,
but accounts for less than 6\% of the V-band luminosity of Phoenix and
0.2\% of the mass. This is thus broadly consistent with the 
comprehensive modelling of the central region of Phoenix based on HST data
by Holtzmann {\it et al.} (2000). They found that star formation 
has been roughly continuous over the lifetime of Phoenix, with no obvious
evidence for strongly episodic star formation, although a mildly varying star
formation rate can fit the data.

Recently in an attempt to probe the optical--HI link further, Gallart 
{\it et al.} (2001) published the first optical radial velocity study of a 
sample of 31 individual stars in Phoenix using VLT/FORS1 in MOS mode. 
Studying the blue end of the optical spectra in the range $\approx$ 
3500$-$6000 \AA \ , which encompasses the Mgb absorption complex, they
determined a mean value of $\vsun = -52\pm6$ km/s.  There is quite a
large offset between this optical measurement and the most likely HI gas
velocity component at $-$23~km/s.  
Gallart {\it et al.} interpret this difference
as either caused by ejection of gas due to supernovae from the most recent 
burst of star formation in Phoenix about 100 Myr ago, or as a consequence of 
ram pressure stripping by a hot intergalactic medium within the Local Group.

The putative association of the HI gas with the optical galaxy is quite 
a crucial measurement, not only because of the direct coupling of the 
(potentially) recently expelled gas, but also because of the unique position 
of Phoenix in the Local Group as potentially (by far) the furthest outlying
satellite of the Galaxy.  At a Galacto-centric distance of $\approx$450 kpc, 
Phoenix could have unprecedented leverage in constraining the mass of the
Galactic Halo out to 450~kpc.  

As part of a long term VLT service programme to measure the radial velocities 
of several outlying Local Group satellites we recently acquired some service 
observations of Phoenix and decided that it was worth re-investigating the 
optical--HI connection with an additional optical velocity measurement.
In this paper we therefore present the results of VLT/FORS1 long slit
observations of Phoenix in the spectral region centred on the Ca~II near
infra-red triplet. The slit position was aligned with 7 stars bright
enough to derive radial velocities from.
Although this is a much smaller number of stars than observed
by Gallart {\it et al.}, each spectrum is of sufficiently high S/N
($\approx$10:1 per continuum \AA) to derive an accurate velocity measurement.

\section{Observations}

The observations were obtained in service-mode, with UT1/FORS1 in the
Long Slit Spectroscopy (LSS) instrument set-up on the nights of 5$-$6 October
2000 and 31 December 2000 to 1 January 2001 (see Table~2).  
The FORS1 long slit spans 6.8 arc-min and was used
with a slit width of 1 arc-sec and the GRIS-600I+15 grism along
with the OG590 order-sorting filter, to cover the Ca~II triplet
wavelength region with as high a resolution as possible.  With this
setting the pixel sampling is close to 1\AA \, per pixel and the
resolution, as estimated from measuring the full width at half maximum of a 
series of unresolved night sky lines, is in the range 3.5--4\AA \,
over the wavelength range 7050$-$9150\AA\,. 
This is the maximum resolution that can be obtained with FORS1
without resorting to a narrower slit. Although this is a wavelength
range at which the FORS1 CCD (Tektronix) has reduced sensitivity it is
where the red giant stars we were aiming to detect are brightest. The
Ca~II triplet is also a useful unblended feature to accurately measure radial
velocities ({\it e.g.} Hargreaves {\it et al.} 1994) and there are
abundant narrow sky lines in this region for wavelength calibration
and/or spectrograph flexure monitoring.

Our observation request consisted of a single slit position, 
or in ESO parlance a single Observation Block (OB).
The data were taken in photometric or thin cirrus conditions with a dark sky 
(moon phase 0.3) and 
generally in sub-arc-second seeing conditions (see Table~2).
The OB was made up of a target observation in conjunction with a
sequence of radial velocity standards (both individual K-giants
and bright globular clusters, see Table~3). The K-giant radial velocity
standards provided the basic velocity standardization in addition to
Ca~II triplet template spectra for comparison with the lower
signal-to-noise dwarf galaxy spectra.  By observing bright globular
clusters with known radial velocities we further sought to test our
methodology with data more closely resembling the primary targets. A
useful by-product of this is additional template spectra to
cross-correlate with the main target objects.

We determined the position of the slit on the galaxy accurately by using
previous FORS1 imaging of this object (Tolstoy {\it et al.} 2000), in
which we tried to hit as many bright giant branch stars, which are likely to be
members of Phoenix, as possible.

\section{Data Reduction and Calibration}

As described in Tolstoy \& Irwin (2000), we did not use the basic
daytime calibrations provided for UT1$+$FORS1 observations to either
flat-field or wavelength calibrate the data. As before we made use of
archive broad I-band filter twilight flat-field frames to create a
master flat-field frame, and we used the plentiful night-sky lines in
this spectral region to wavelength calibrate the data.  The
two-dimensional non-linear wavelength distortion of the long-slit data
was again found to be stable enabling us to make use of the same
two-dimensional wavelength calibration of all the data for a given
night.  This allows us to correct the distortion of the radial
velocity standards, which lack significant sky-lines due to their short
exposure times.  After extracting all the spectra, cross-correlation
of either contemporaneously extracted sky spectra or atmospheric
absorption features in the object spectra were used to correct for
shifts in the zero-point of the wavelength calibration caused by
spectrograph flexure or mis-centering of objects in the slit.  All of
our calibration and data reduction was carried out using the IRAF
package.
\footnote{IRAF is distributed by the National Optical Astronomy
Observatories, which are operated by the Association of Universities
for Research in Astronomy, Inc., under cooperative agreement with the
National Science Foundation.}

The observation of the Phoenix dwarf spheroidal galaxy consisted of
3$\times$1000s exposures.  There were no measurable internal shifts
found for frames within a sequence, so these were averaged with
k-sigma clipping used to eliminate cosmic ray events.  
Full two-dimensional wavelength calibration, using the night sky lines,
was then done on the combined data.
The globular cluster observation was an integration of 250s
which was treated in the same way as the target galaxy
observations.

The position of the slit across the central region of Phoenix, taken
from the FIMS file used to prepare the OBs, is shown in Figure~1. Also
marked are the 7 stars clearly detected in the Phoenix spectra. The 
position of all the resolved objects detected in the Phoenix long-slit spectra
are plotted on the Colour-Magnitude Diagram (CMD) of Phoenix (from Tolstoy 
{\it et al.} 2000) in Figure~2.  From their location on the CMD, and also 
position on
the slit, virtually all of the stars are high probability members of Phoenix.
Three stars (1058, 899 and 496) are likely to be red giant branch (RGB) 
members, star 845 is probably on the asymptotic giant branch (AGB), 
whereas 1274 and 1328 could be evolved blue loop main sequence stars. 
Howver, 1328 and 1274 (and also 773) lie in  
crowded regions, and could easily be RGB stars with  
contaminating flux from a nearby object affecting their photometry.
The identification of all the 
resolved objects in the slit and their location on the CMD provides an 
excellent independent corroboration of the optical velocity determinations.
However, we also note that stars 1328, 773 and 1274 also lie in a region of 
the CMD where there is a higher probability of foreground contamination.

With the usual caveats of over interpretation of posterior statistics
we can make an estimate of the likelihood of foreground contamination
as follows.  Phoenix is at very high Galactic latitude, b=$-$68.9,
which together with a Galactic longitude of l=272.2 suggests that the 
contamination by foreground stars in likely to be very low.
The radial velocity distribution of potential Galactic foreground
contaminants comes from K and M dwarfs in the thin disk, thick disk and
halo.  In this direction the velocity dispersion of foreground stars is
dominated by the $\sigma_W$ component which has a value of 
$\approx$20km/s, 40km/s and 100km/s for thin disk, thick disk and halo stars 
respectively (eg. Carney et al. 1989; Mihalas \& Binney 1981).
The heliocentric velocity offset of the foreground components due to the 
relative solar motion is $\approx$11km/s, 25km/s and 90km/s respectively.
All of these populations are potential contaminants but the positive
systemic velocities (and to some extent the large velocity dispersions)
mitigate against significant pollution.  Further constraints come from
the spatial density.  The Phoenix CMD was constructed over a
region 7$\times$7 arc-min in size and over the total magnitude and
colour range of the spectroscopic observations (18.5 $<$ R $<$ 21 and
1.0 $<$ B$-$R $<$ 3.0), has 380 stellar objects detected.  Of these
approximately 90$-$95\% are expected to be stars in Phoenix (e.g.
Ratnatunga \& Bahcall 1985), leaving some 20$-$30 stars as potential foreground
contamination.  The long-slit used in the FORS1 observations covers
0.2\% of this area therefore the expected number of random contaminating
foreground stars in the slit is less than 0.1 over this magnitude and colour
range.  However, given that we did some prior selection of brighter stars 
in the above magnitude colour range by aligning the slit to include three 
relatively bright objects, it is possible that 1 or more of these selected 
brighter objects could be a foreground object.
Since the expected heliocentric radial velocity 
of Phoenix is low, contamination by say a single disk foreground K/M dwarf 
(the most likely contaminant) will be both difficult to spot but conversely
would also have relatively little impact on the derived systemic velocity.
We disuss this issue further in section 5.

\section{Extracting and Wavelength Calibrating the Spectra}

For the globular clusters and dSphs, image sections summed along
the spatial axis were used to identify resolved stellar, or background
galaxy images, in the slit and also to define the underlying
unresolved background of the dwarf spheroidal galaxy for spectroscopic
extraction.  Spectra for these objects plus the integrated background
were then extracted in the usual manner taking care to optimise the
region chosen to represent the local sky.  The computed sky spectrum
was also extracted contemporaneously for later use in checking the wavelength 
solution systematics (see Figure~3).  
For the radial velocity standards and also the target objects in the long slit,
contributions to systematics from slit centering is also an issue
that limits the achieved final accuracy.  This is reflected in the scatter 
($\sigma_v \sim 5$km/s) of the derived velocities for the standards in Table~4
which is dominated by slit-centering errors and limited by the accuracy of 
post-reduction corrections based on atmospheric absorption lines.
The details are identical to those described in Tolstoy \& Irwin (2000).

Wavelength calibration was primarily based on one of the central sky spectra 
taken from the Phoenix data.  All of the remaining extracted one-dimensional 
sky spectra for cluster calibration and target objects were then 
cross-correlated against the reference image sky spectrum, to adjust the 
zero-point of the wavelength solution.  This enables us to monitor the 
stability of the wavelength solution and hence determine and track any
flexure of the spectrograph.  The cross-correlation of sky lines within a 
single two-dimensional frame gives negligible velocity shifts of 
$\pm 1-$2 km/s which is comparable to the juxtaposition of the
wavelength calibration errors and correlation-function estimation errors.
Sky lines are generally not visible in the standard star spectra so the 
wavelength zero-point here was checked using the main atmospheric absorption
A-band.

Since we had previous observations taken with a similar setup (Tolstoy \&
Irwin 2000) we could make further use of our earlier cluster and standard star
observations to standardise the velocity determinations.  After applying all
the various checks on the wavelength calibration we estimate that the final
systematics of the wavelength solution are in the range 5$-$10~km/s, and are
mainly dominated by corrections for slit-centering errors.  These are
comparable to the {\it rms} errors, dominated by photon noise, in the
cross-correlation of the fainter target objects, but are more than adequate 
for the current goal.

\section{Determining the Radial Velocities}

The main advantages of using the Ca~II triplet region for radial velocity
determination are three-fold: the relative brightness and uniformity of the 
continuum flux from K-giants; the narrowness $\approx$3\AA \ of the three
Gaussian-like Ca~II absorption features; and the abundance of narrow night sky
lines for wavelength calibration.  Although at first sight the latter might
seem a significant drawback too, in practice the velocity template spectrum
is only ``active'' over the narrow regions of $\approx$10\AA \ centred on 
each of the triplet lines.  This minimises the influence of the, on average,
relatively bright sky and means that in practice it is feasible to measure
reliable cross-correlations several magnitudes below the average sky level
in this region.

For all the globular cluster and dSph observations, individual resolved 
objects were extracted separately in an attempt to identify system members 
and possible foreground star, or even background galaxy, interlopers.  
Radial velocities were measured for all the spectra, as previously
described by Tolstoy \& Irwin (2000), using the standard Fourier 
cross-correlation package in IRAF, FXCOR, after suitable continuum removal 
and apodizing of the target objects.  The template function consisted of the
spectrum of a bright radial velocity standard with the continuum fitted and
removed to give a mean level of zero for the continuum.  All regions outside
of the Ca~II triplet lines in the template were then identically set to zero, 
leaving ``active'' only the narrow region of $\approx$10\AA \ centred on 
each of the three absorption components.

An example of the results for Phoenix is shown in Figure~4 where the Ca~II 
triplet is clearly visible in the target spectrum.  The measured radial 
velocities are then corrected
for: the template radial velocity offset; the topo-centric correction to a 
helio-centric system; and for flexure/slit mis-centering.  The final results
for the individual objects extracted from the spectra are listed 
in Table~4, including Phoenix, the radial velocity standards and the globular 
cluster Pal~12. 

Virtually all of the extracted spectra for Phoenix have radial velocities 
that are consistent with membership given the likely (unknown) optical 
velocity dispersion of $\approx$10 km/s.  The range of observed velocities
in the 7 stars is from $-$31 km/s to $+$9 km/s 
with a mean of $-$13.4 km/s, which
compares favourably with the HI velocity of $-$23 km/s.  The error in the 
estimate is dominated by a combination of systematic residuals, random
photon noise errors and the velocity dispersion of the dwarf.  Combining
these error estimates leads to a standard error in the derived mean
velocity of around $\pm$9 km/s.  It is interesting to note that the
formal $rms$ dispersion about the mean for the 7 stars is $\sigma = 14.2$km/s.
Therefore purely on kinematic grounds, it is difficult to rule out a 
modest disk contamination since the expected mean radial velocity
for disk stars is $+$11km/s in this direction, with dispersion of $\pm$20km/s.
However, as noted earlier, the converse is also true, in that contamination by
one or two disk stars has little impact on the derived systemic velocity
of the dwarf. 
As noted in section 3. the most likely foreground contaminants are the 3 
relatively bright stars (773,496,1274) used for selecting the most 
favourable slit location.  Of these, 496, lies on the RGB and is
therefore unlikely to be foreground.  In the worst case scenario, if the 
other two stars are foreground, excluding them shifts the derived optical 
radial velocity by +3 km/s, which is a negligible change relative to the
inherent measuring errors of $\pm$9km/s.

\section{Discussion}

Our determination of the optical velocity for Phoenix of $-$13km/s $\pm$9km/s
closely matches the most likely HI velocity for this galaxy of $-$23 km/s 
derived by St-G\'{e}rmain {\it et al.} (1999).  In a recent review of the impact of
the VLT on Local Group Galaxies, Held (2001) quotes recent UVES measurements
of giant stars in the Phoenix galaxy that agree to within a few km/s of the 
neutral gas velocity.  It is difficult to directly reconcile these much smaller
negative optical radial velocities with the $-$52km/s $\pm$6 km/s recently
determined by Gallart {\it et al.} (2001).

It is highly unlikely that Galactic foreground contamination could have 
affected any of the optical results significantly and the HI gas velocity
result of $-$23km/s seems convincingly secure as argued by 
St-G\'{e}rmain {\it et al.} 
(1999).  There is also unlikely to be contamination of our results from the 
two main tidal streams enveloping our Galaxy: the Magellanic stream and the 
Sagittarius Dwarf tidal stream.  The Magellanic Stream is predominantly gas
with no unambiguous detection of a stellar component yet reported.  Therefore,
while the Magellanic Stream can contaminate the gas distribution in the Phoenix
direction, it is unlikely to contribute to contamination of the optical 
velocities.  Likewise, pollution by the Sagittarius Dwarf tidal stream is also
highly unlikely since the projection of its current orbit does not pass close 
to the line-of-sight to Phoenix.

The gradient in the distribution of young stars in Phoenix (e.g.,
Ortolani \& Gratton 1988; Mart\'inez-Delgado {\it et al.} 1999)
suggests that the recent star formation has been moving from east to
west across the central component of Phoenix in a manner consistent
with self-propagating star formation theories. The youngest stars are
thus spatially overlapping the position of the HI cloud at $-$23km/s (as
originally pointed out by Young \& Lo 1997).  The relative position and
velocity of this cloud combined with the evidence of recent star
formation provides evidence that a burst of star formation can disrupt
and potentially blow out gas from the centre of a dwarf galaxy,
inhibiting further star formation (e.g., Dekel \& Silk 1986; Mac Low \&
Ferrara 1999), although it is not the only possible explanation for
what is seen (cf. ram pressure stripping scenarios as in Gallart {\it et al.}
2001).

However, we can conclude that our results in conjunction with those of Held 
(2001) unequivocally show that modest amounts, 
($\approx 10^5 \Msun$), of HI gas are associated with the Phoenix dwarf 
galaxy albeit somewhat offset from the optical centre of the galaxy.  

\acknowledgements{{\bf Acknowledgments:} 
These data were taken in service mode, and we thank
the Paranal Science Operations Staff for their efforts.
ET gratefully acknowledges support from a fellowship of the
Royal Netherlands Academy of Arts and Sciences and support from a Visitor 
Grant at the Institute of Astronomy, Cambridge. }

\newpage

\def\x{\enspace}
\def\xx{\enspace\enspace}
\def\xxx{\enspace\enspace\enspace}
\def\xxxx{\enspace\enspace\enspace\enspace}
\def\xxxxx{\enspace\enspace\enspace\enspace\enspace}
\def\jref#1 #2 #3 #4 {{\par\noindent \hangindent=3em \hangafter=1
      \advance \rightskip by 5em #1, {\it#2}, {\bf#3}, {#4} \par}}
\def\ref#1{{\par\noindent \hangindent=3em \hangafter=1
      \advance \rightskip by 5em #1 \par}}
\def\endtable{\endgroup}
\def\tableheight{\vrule width 0pt height 8.5pt depth 3.5pt}
{\catcode`|=\active \catcode`&=\active
    \gdef\tabledelim{\catcode`|=\active \let|=\vbar
                     \catcode`&=\active \let&=\nobar} }
\def\table{\begingroup
    \def\twidth{\hsize}
    \def\tablewidth##1{\def\twidth{##1}}
    \def\defaultheight{\vrule width 0pt height 8.5pt depth 3.5pt}
    \def\heightdepth##1{\dimen0=##1
        \ifdim\dimen0>5pt
            \divide\dimen0 by 2 \advance\dimen0 by 2.5pt
            \dimen1=\dimen0 \advance\dimen1 by -5pt
            \vrule width 0pt height \the\dimen0  depth \the\dimen1
        \else  \divide\dimen0 by 2
            \vrule width 0pt height \the\dimen0  depth \the\dimen0 \fi}
    \def\spacing##1{\def\defaultheight{\heightdepth{##1}}}
    \def\nextheight##1{\noalign{\gdef\tableheight{\heightdepth{##1}}}}
    \def\end{\cr\noalign{\gdef\tableheight{\defaultheight}}}
    \def\zerowidth##1{\omit\hidewidth ##1 \hidewidth}
    \def\hline{\noalign{\hrule}}
    \def\skip##1{\noalign{\vskip##1}}
    \def\bskip##1{\noalign{\hbox to \twidth{\vrule height##1 depth 0pt \hfil
        \vrule height##1 depth 0pt}}}
    \def\header##1{\noalign{\hbox to \twidth{\hfil ##1 \unskip\hfil}}}
    \def\bheader##1{\noalign{\hbox to \twidth{\vrule\hfil ##1
        \unskip\hfil\vrule}}}
    \def\spanloop{\span\omit \advance\mscount by -1}
    \def\extend##1##2{\omit
        \mscount=##1 \multiply\mscount by 2 \advance\mscount by -1
        \loop\ifnum\mscount>1 \spanloop\repeat \ \hfil ##2 \unskip\hfil}
    \def\vbar{&\vrule&}
    \def\nobar{&&}
    \def\hdash##1{ \noalign{ \relax \gdef\tableheight{\heightdepth{0pt}}
        \toks0={} \count0=1 \count1=0 \putout##1\end
        \toks0=\expandafter{\the\toks0 &\end} \xdef\piggy{\the\toks0} }
        \piggy}
    \let\e=\expandafter
    \def\putspace{\ifnum\count0>1 \advance\count0 by -1
        \toks0=\e\e\e{\the\e\toks0\e&\e\multispan\e{\the\count0}\hfill}
        \fi \count0=0 }
     \def\putrule{\ifnum\count1>0 \advance\count1 by 1
        \toks0=\e\e\e{\the\e\toks0\e&\e\multispan\e{\the\count1}\leaders\hrule\
hfill}
        \fi \count1=0 }
    \def\putout##1{\ifx##1\end \putspace \putrule \let\next=\relax
        \else \let\next=\putout
            \ifx##1- \advance\count1 by 2 \putspace
            \else    \advance\count0 by 2 \putrule \fi \fi \next}   }
\def\tablespec#1{
    \def\vdimens{\noexpand\tableheight}
    \def\tabby{\tabskip=0pt plus100pt minus100pt}
    \def\r{&################\tabby&\hfil################\unskip}
    \def\c{&################\tabby&\hfil################\unskip\hfil}
    \def\l{&################\tabby&################\unskip\hfil}
    \edef\templ{\noexpand\vdimens ########\unskip  #1
         \unskip&########\tabskip=0pt&########\cr}
    \tabledelim
    \edef\body##1{ \vbox{
        \tabskip=0pt \offinterlineskip
        \halign to \twidth {\templ ##1}}} }

\centerline{\bf Table 1: Phoenix}
\vskip-1cm
$$
\table
\tablespec{\l\c\c\c\l\l}
\body{
\skip{0.06cm}
\hline
\skip{0.025cm}
\hline
\skip{.2cm}
& Object  &  RA (J2000) Dec      & Distance & M$_V$ & type & ref &\end
&         &                      & ~~(kpc)  &       &      &     &\end
\skip{.1cm}
\hline
\skip{.05cm}
\hline
\skip{.45cm}
& Phoenix & 01 51 06 $-$44 26 42 & 445 $\pm$ 30 & $-$10.1 & dIrr/dSph & Ortolani \& Gratton 1988 &\end
&         &                      &              &         &           & van de Rydt {\it et al.} 1991 &\end
\skip{.25cm}
\hline
\skip{0.025cm}
\hline
}
\endtable
$$

\vskip1cm

\centerline{\bf Table 2: The Observations}
\vskip-1cm
$$
\table
\tablespec{\l\l\l\r\c\c\l}
\body{
\skip{0.06cm}
\hline
\skip{0.025cm}
\hline
\skip{.2cm}
& Date & Begin &Object & Exptime & Airmass & Seeing & Comments &\end
& & UT & & ~~(secs) & & ~(arcsec) &&\end
\skip{.1cm}
\hline
\skip{.05cm}
\hline
\skip{.45cm}
& 06 Oct 2000 & 00:33 & HD~203638  & 1    & 1.01  &0.85 &photometric&\end
&             & 00:46 & Pal~12     & 250  & 1.02  &0.60 &           &\end
\skip{.1cm}
& 01 Jan 2001 & 00:34 & HD~8779    & 1    & 1.15  &1.02 &thin cirrus&\end
&             & 01:00 & HD~8779    & 1    & 1.20  &0.95 &           &\end
&             & 01:12 & Phoenix--1 & 1000 & 1.13  &1.13 &variable seeing &\end
&             & 01:30 & Phoenix--2 & 1000 & 1.14  &0.97 &           &\end
&             & 01:48 & Phoenix--3 & 1000 & 1.20  &1.03 &           &\end
&             & 02:11 & HD~8779    & 1    & 1.46  &0.79 &           &\end
\skip{.25cm}
\hline
\skip{0.025cm}
\hline
}
\endtable
$$
\vskip1cm

\renewcommand{\thefootnote}{\fnsymbol{footnote}}

\centerline{\bf Table 3: The Calibrators}
\vskip-1cm
$$
\table
\tablespec{\l\l\c\r\l}
\body{
\skip{0.06cm}
\hline
\skip{0.025cm}
\hline
\skip{.2cm}
& Object & Class &V & v$_{\odot}$ &\end
&  & && ~~km/s & Ref &\end
\skip{.1cm}
\hline
\skip{.05cm}
\hline
\skip{.45cm}
& Pal~12      & Cluster & 11.99 & +27.8 & Harris 1996&\end
\skip{.2cm}
\skip{.2cm}
& HD~203638& K0 III & 5.77 & $+$21.9 &&\end
\skip{.1cm}
& HD~8779  & gK0    & 6.41 & $-$5.0  &&\end
\skip{.1cm}
& HD~107328& K1 III & 4.96 & $+$35.7 &RV calibrator &\end
\skip{.25cm}
\hline
\skip{0.025cm}
\hline
}
\endtable
$$

\newpage

\centerline{\bf Table 4: Individual Velocity Results}
\vskip-1cm
$$
\table
\tablespec{\l\r\r\l}
\body{
\skip{0.06cm}
\hline
\skip{0.025cm}
\hline
\skip{.2cm}
& Object         & $\vsun$  & $rms$ error & Comment &\end
&                & ~~(km/s) & ~~(km/s)    &         &\end
\skip{.1cm}
\hline
\skip{.05cm}
\hline
\skip{.45cm}
& HD~203638       & $+$28.6  & $\pm 2.5$   & Systematic errors&\end
\skip{.1cm}
& HD~8779--1      & $-$8.9   & $\pm 3.0$   & due to flexure and&\end
& HD~8779--2      & $+$2.7   & $\pm 2.4$   & slit centering are &\end
& HD~8779--3      & $+$3.7   & $\pm 2.5$   & typically $\pm$ 5--10 km/s &\end
\skip{.2cm}
& Pal~12--1       & $+$26.6  & $\pm 2.2$   &         &\end
& Pal~12--2       & $+$26.1  & $\pm 3.8$   &         &\end
& Pal~12--3       & $+$28.8  & $\pm 3.9$   &         &\end
& Pal~12--4       & $-$24.9  & $\pm 5.2$   & Probably a non-member &\end
\skip{.2cm}
& Phoenix--773   & $-$8.9   & $\pm 3.0$   & The most likely   &\end
& Phoenix--496   & $-$22.4  & $\pm 2.5$   & systemic HI velocity is &\end 
& Phoenix--845   & $-$26.8  & $\pm 3.5$   & $\vsun = -23$ km/s &\end
& Phoenix--1058  & $-$3.6   & $\pm 3.9$   &         &\end
& Phoenix--899   & $+$8.8   & $\pm 5.8$   &         &\end
& Phoenix--1274  & $-$31.4  & $\pm 11.0$  &         &\end
& Phoenix--1328  & $-$9.3  & $\pm 10.0$   &         &\end
\skip{.1cm}
\hline
\skip{.05cm}
\hline
}
\endtable
$$
The radial velocity primary reference standard used was based on previous
observations of HD~107328 from Tolstoy \& Irwin (2000).  \newline
The rms errors were estimated directly from the centroiding of a
general Gaussian fit to the primary cross-correlation peak. \newline
Continuum signal:to:noise was typically in the range 5--10 per pixel
in the Ca~II triplet region.

\clearpage

\begin{figure}
\caption{The central 4 arcmin of the Phoenix dwarf galaxy
with the adopted slit position marked. North is up and East to the left.
This diagram comes from the FIMS obervation preparation tool output. The 
image is a 600sec B-band observation made with FORS1 in August 1999.
Marked on the image are the 7 stars identified in our longslit spectrum for 
which reliable individual spectra could be usefully extracted. 
}
\label{first}
\end{figure}
\newpage

\begin{figure}
\caption{Phoenix (R, B$-$R) Colour-Magnitude Diagram from VLT data taken in
August 1999 (Tolstoy {\it et al.} 2000). Plotted as star symbols are objects
in the slit for which we could extract individual spectra.
The labels correspond to those in Figure~1 and also in Table~4. }
\label{second}
\end{figure}
\newpage
\clearpage

\begin{figure}
\caption{The spectrum of star \#1058 (in observed counts) compared with the 
spectrum of one of the observations of the radial velocity standard HD8779 
(scaled down by a factor of 250 and shifted vertically for ease of reference).
Also plotted for comparison is a sky spectrum (scaled down by a factor 
of 40) illustrating the potential problems of residual sky line contamination 
on the spectrum of the target.  A 3-point linear box-car filter was used to 
smooth the template and target spectra for plotting purposes.  The example 
night sky spectrum is unsmoothed.  Despite the brightness of the night sky 
lines, the extracted target spectrum  has a clearly visible Ca~II triplet.}
\label{third}
\end{figure}
\newpage
\begin{figure}
\caption{The normalised cross-correlation of the Phoenix spectrum for star 
\#1058 shown in Figure~3, against a template spectrum constructed from
an observation of HD8779.  Prior to cross-correlation, both spectra are 
continuum-corrected and then the template spectrum is set to zero for all 
regions outside of the $\approx$10\AA \ zone for each of the CaII absorption 
lines.  No template, flexure or helio-centric corrections 
have been made for this plot. }
\label{fourth}
\end{figure}


\begin{references}

Carignan, C., Demers, S. \& C\^{o}te, S. 1991 ApJL, 381, 13

Carney, B.W., Latham, D.W., \& Laird, J.B. 1989 AJ, 97, 423

Dekel, A. \& Silk, J. 1986 ApJ, 303, 39

Gallart, C., Mart\'inez-Delgado, D., G\'omez-Flechoso, M.A. \& Mateo, M.,
2001 AJ, 121, 2572

Hargreaves, J.C., Gilmore, G., Irwin, M.J. \& Carter, D. 1994 MNRAS, 269, 957

Held, E.V., Saviane, I. \& Momany, Y. 1999 A\&A, 345, 747

Held, E.V. 2001 in press, Proceedings of the ESO Workshop Scientific 
Drivers for ESO Future VLT/VLTI Instrumentation, eds, J. Bergeron \& 
G. Monnet {\it astro-ph/0109548}

Holtzmann, J.A., Smith, G.H. \& Grillmair, C. 2000 AJ, 120, 3060

Mac Low, M.-M. \& Ferrara, A. 1999 ApJ, 513, 142

Mart\'inez-Delgado, D., Gallart, C. \& Aparicio, A. 1999 AJ, 118, 862

Mihalas, D., \& Binney, J., in {\it Galactic Astronomy}, 
W.H. Freeman and Company.

Oosterloo, T., Da Costa, G.S. \& Staveley-Smith, L. 1996 AJ, 112, 1969

Ortolani, S. \& Gratton, R.G. 1988 PASP, 100, 1405

Ratnatunga, K. \& Bahcall, J.N. 1985 ApJS 59, 63

St-G\'{e}rmain, J., Carignan, C., C\^{o}te, S., Oosterloo, T. 1999 AJ, 118, 1235

Tolstoy, E., Gallagher, J.S., Greggio, L., Tosi, M., De Marchi, G., 
Romaniello, M., Minniti, D. \& Zijlstra, A.A. 2000, ESO Messenger, 99, 16

Tolstoy, E. \& Irwin, M.J., 2000 MNRAS, 318, 1241

van de Rydt, F., Demers, S. \& Kunkel, W.E. 1991 AJ, 102, 130

Young, L.M., \& Lo, K.Y. 1997 ApJ, 490, 710

\end{references}
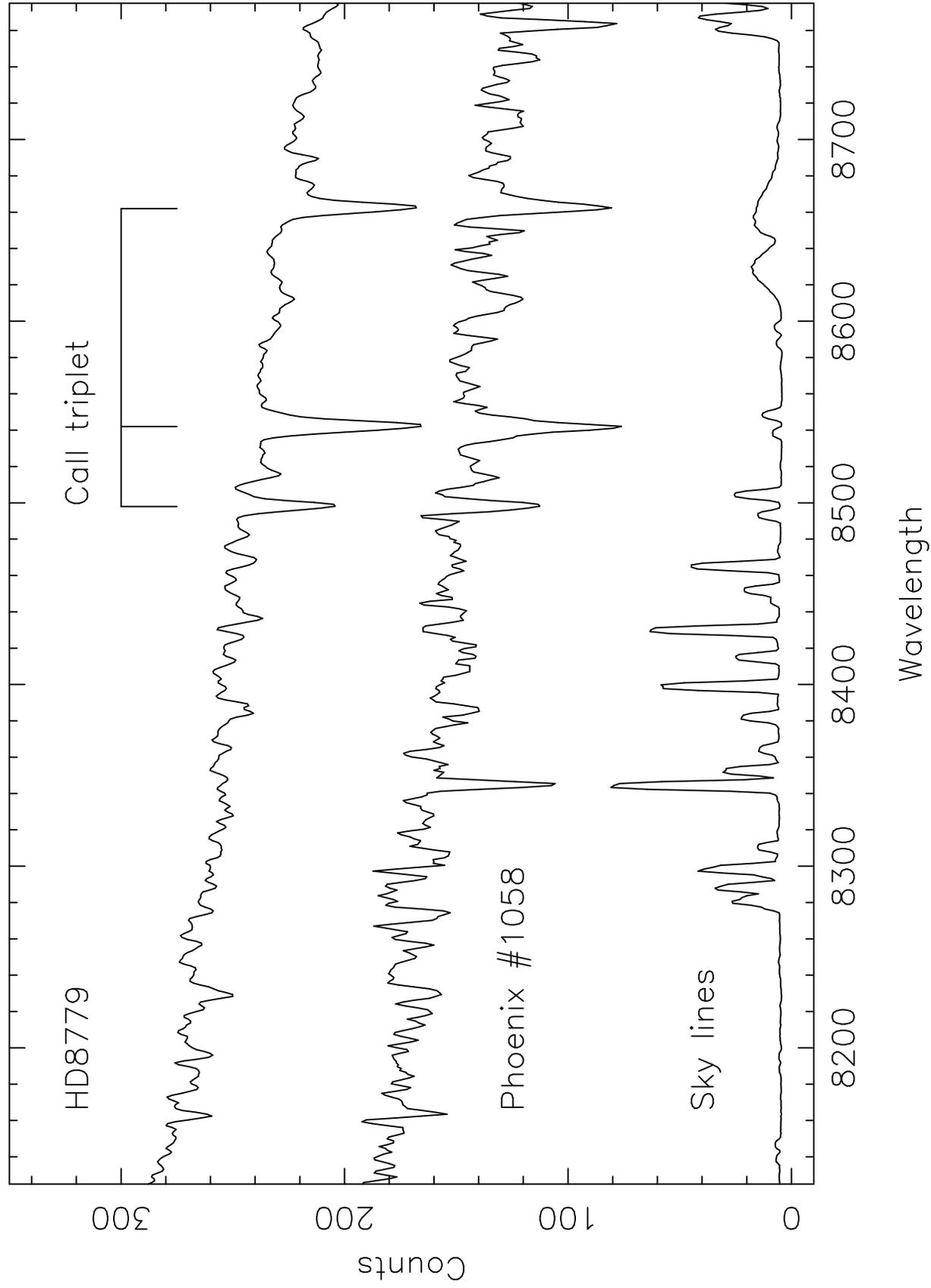
\end{document}